\documentclass[10pt]{article}

\usepackage{amsmath}
\usepackage{amssymb}
\usepackage{amsfonts}
\usepackage{graphicx}
\usepackage{lscape}
\usepackage{mathrsfs}
\usepackage{multirow}
\usepackage{rotating}
\usepackage{wrapfig}
\usepackage{subfigure}
\usepackage{url}
\usepackage{bm}
\usepackage{ulem}
\usepackage{enumitem}

\def\aa{A\&A}

\def\aj{AJ}
\def\apj{ApJ}

\begin{document}

\setcounter{figure}{0}
\setcounter{table}{0}
\setcounter{footnote}{0}
\setcounter{equation}{0}

\vspace*{0.5cm}

\noindent {\Large THE DEFLECTION OF LIGHT INDUCED BY THE SUN'S GRAVITATIONAL FIELD AND MEASURED WITH GEODETIC VLBI}
\vspace*{0.7cm}

\noindent\hspace*{1.5cm} O.A. TITOV$^1$, A.A. GIRDIUK$^2$\\
\noindent\hspace*{1.5cm} $^1$ Geoscience Australia\\
\noindent\hspace*{1.5cm} PO Box 378, Canberra, ACT 2601, Australia\\
\noindent\hspace*{1.5cm} e-mail: oleg.titov@ga.gov.au\\
\noindent\hspace*{1.5cm} $^2$ Institute of Applied Astronomy, Russian Academy of Sciences\\
\noindent\hspace*{1.5cm} Kutuzov Quay 10, 191187, Saint-Petersburg, Russia\\
\noindent\hspace*{1.5cm} e-mail: girduik@ipa.nw.ru\\

\vspace*{0.5cm}

\noindent {\large ABSTRACT.} 

The Sun's gravitational field deflects the apparent positions of close objects in accordance with the formulae of general relativity. 
Optical astrometry is used to test the prediction, but only with the stars close to the Sun and only during total Solar eclipses. 
Geodetic Very Long Baseline Interferometry (VLBI) is capable of measuring the deflection of the light from distant radio sources anytime and across the whole sky. 
We show that the effect of light deflection is equivalent to the gravitational delay calculated during the reduction of VLBI data. 
All reference radio sources display an annual circular motion with the magnitude proportional to their ecliptic latitude. 
In particular, radio sources near the ecliptic pole draw an annual circle with magnitude of 4 mas. 
This effect could be easily measured with the current precision of the geodetic VLBI data.

\vspace*{1cm}

\noindent {\large 1. INTRODUCTION}

\smallskip

Very Long Baseline Interferometry (VLBI) is capable of measuring precise
group delays - the difference in arrival times of radio waves at two
radio telescopes (Schuh \& Behrend, 2012) from distant extragalction radio sources (quasars).
Accurate positions of these radio sources are obtained with an accuracy of 40
microarcsec (Ma et al., 2009). 

Gravitational time delay caused by the Solar gravitational field is calculated during the reduction of geodetic VLBI data
(Shapiro, 1964, 1967). 
The conventional formula for calculating gravitational delay is formulated
in terms of the positions of the radio telescopes within the
barycentric reference frame of the Solar System (Kopeikin, 1990; Eubanks et al., 1991; Soffel et al., 1991; Klioner, 1991), 
rather than the baseline length between the radio telescopes. 

We propose an alternate gravitational delay formula using a Taylor
series expansion. We show that the conventional formula can be split into a sum of several terms, and the major term
links the gravitational delay and the well-known formula for the light deflection angle. 
The light deflection angle can be considered equivalent for all baselines and estimated for each radio source at times of interest. We develop a new approach to probe the formula for the 
light deflection angle at an arbitrary elongation from the Sun. Finally, estimates of the light deflection for several reference radio sources
based on VLBI observations in 1991-2001 are presented.

\vspace*{0.7cm}

\noindent {\large 2. GRAVITATIONAL DELAY VS LIGHT DEFLECTION}

\smallskip

The light deflection angle $\alpha$ at an arbitrary elongation $\theta$ from the Sun, is given by

\begin{equation}\label{alpha}
\alpha =  \frac{(\gamma + 1)GM}{c^{2}r}\frac{\sin\theta}{1-\cos\theta}
\end{equation}

where $G$ is the gravitational constant, $M$ is the mass of a gravitational body, $c$ is the speed of light,
$r$ is the distance from the Earth to the Sun, {$\gamma$} is the parameter of the Parametrised Post-Newtonian formalism (PPN),
$\theta$ is the elongation angle (i.e. the angular distance between the Sun and the gravitational body, as measured by the Earth observer) (Shapiro, 1967; Ward, 1970).
The conventional gravitational delay is calculated as follows
\begin{equation}\label{shapiro}
\tau_{grav}=\frac{(\gamma+1)GM}{c^3}\ln \frac{|\vec{r_1}|+(\vec{s}\cdot\vec{r_1})}{|\vec{r_2}|+(\vec{s}\cdot\vec{r_2})},
\end{equation}
The VLBI total delay model also comprises a term due to the transformation from the barycentric to the geocentric reference frames.
\begin{equation}\label{t_geom}
\tau_{coord} = \frac{(\gamma+1)GM}{c^2r}\frac{(\vec{b}\cdot\vec{s})}{c}
\end{equation}
It can be shown that the three formulae are approximately linked as (Titov \& Girdiuk, 2015)
\begin{equation}\label{t_geom}
\tau_{GR} \approx \tau_{grav} + \tau_{coord} = \alpha \frac{b}{c}\sin\varphi\cos A
\end{equation}
where angles $\varphi, \theta, A$ are linked by the standard spherical triangle formula
\begin{equation}\label{psi_angl}
	\begin{array}{c}
\cos\psi=-\cos\varphi\cos\theta-\sin\varphi\sin\theta\cos A,
	\end{array}
\end{equation}
Converting the three dot products with angles $\varphi,\; \psi, \; \theta$ (Fig 1), where 
\begin{equation}\label{angl_mult}
(\vec{b}\cdot\vec{s})=|\vec{b}|\cos\varphi,\;\;\; (\vec{b}\cdot\vec{r_2})=|\vec{b}||\vec{r_2}|\cos\psi,\;\;\; (\vec{r_2}\cdot\vec{s})=-|\vec{r_2}|\cos\theta
\end{equation}
\vspace{-0.9cm}
\begin{figure}[h!]
	\begin{minipage}[h]{0.49\linewidth}
		\begin{center}
			\includegraphics[width=0.45\linewidth]{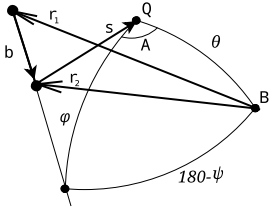}
		\end{center}
	\end{minipage}
	\hfill
	\begin{minipage}[h]{0.49\linewidth}
		\caption{Angles $\varphi$, $\psi$, $\theta$ and $A$, originated by position of gravitational mass (B), quasar (Q) and baseline vector ($\vec{b}$). If the Sun plays the role of the gravitational mass, then, the point B on Fig 1 is also the position of the Solar system barycentre.}\label{one}
	\end{minipage}
\end{figure}

More accurately, the total contribution of the post-Newtonian (PN) general relativity effect to the total VLBI delay is given by (we recall now $\gamma = 1$ for the case of general relativity)
\begin{equation}\label{total}
		\begin{array}{ll}
\tau_{GR} \approx&\frac{2GMb}{r_2c^{3}}\frac{\sin\varphi\sin\theta\cos A}{1-\cos\theta}+\frac{GM}{c^{3}}\frac{b^2(1-\cos^2\varphi\cos^2\theta)}{r_2^2(1-\cos\theta)} -\frac{GM}{c^{3}}\frac{b^2\sin^2\varphi\sin^2\theta\cos^2 A}{r_2^2(1-\cos\theta)^2} 
		\end{array}
\end{equation}
The two additional terms appear because the elongation angles $\theta_{1}$ and $\theta_{2}$ between the direction to the observed radio source and the centre of the gravitational body, as measure from each radio telescope, are not equivalent. The parallactic effect for a baseline of 6,000 km is about 8 arcsec, and the deflection angles $\alpha_{1}$ and $\alpha_{2}$ for the "first" and the "second" radio telescope differing. At a large elongation the additional terms are negligible, so in the small angle approximation this effect should be considered.
\begin{equation}\label{Einstein}
		\begin{array}{ll}
\tau_{GR} & = \frac{4GM}{c^{2}R_{2}}\frac{b}{c}\sin\varphi(\cos A - \frac{b}{2R_{2}}\sin\varphi\cos 2A) = \alpha_{2} \cdot \frac{b}{c}\sin\varphi(\cos A - \frac{b}{2R_{2}}\sin\varphi\cos 2A)
		\end{array}
\end{equation}
where $\alpha_{2}$ is the classical deflection angle for light propagated through a gravitation field (Einstein, 1916) for the second station $\alpha_{2} =  \frac{4GM}{c^{2}R_{2}}$ and $R_{2} = r_{2}\cos\theta_{2}$.
In an alternative form 
\begin{equation}\label{Einstein}
\tau_{GR} = (\alpha_{2} - \frac{GM}{c^{2}R_{2}}\frac{b}{R_{2}}\frac{\sin\varphi\cos 2A}{\cos A}) \cdot \frac{b}{c}\sin\varphi\cos A =  (\alpha_{2} + \alpha'_{2}) \cdot \frac{b}{c}\sin\varphi\cos A 
\end{equation}
where $\alpha'_{2}$ is the additional deflection angle $\alpha'_{2} =  \frac{GM}{c^{2}R_{2}}\frac{b}{R_{2}}\frac{\sin\varphi\cos 2A}{\cos A}$

\vspace*{0.7cm}

\noindent {\large 3. OBSERVATIONAL DATA}

We processed the available VLBI data between 1991 and 2001 using the OCCAM software (Titov et al., 2004). A fraction of well-established reference radio sources were assigned as astrometrically unstable, i.e. their coordinates were not fixed. The observational group delays were approximated by the theoretical values, and the O--C (observed minus calculated) differences were used to obtain a set of estimates of the standard daily parameters (Earth orientation, station positions, wet troposphere delays and gradients, etc.) and corrections to the selected radio source coordinates (right ascension and declination) regardless of its elongation from the Sun. The consensus gravitational delay (3) and the coordinate term (4) were not applied for the calibration of the group delay for the radio sources. The ionosphere fluctuations were calibrated in the conventional way, and, as it will be shown later, this makes possible VLBI observations of radio sources in the range of 1$^\circ$.5 to 3$^\circ$ from the Sun.
\begin{figure}[h]
	\begin{minipage}[h]{0.49\linewidth}
		\begin{center}
			\includegraphics[scale=0.5, trim= 60 270 80 260,clip]{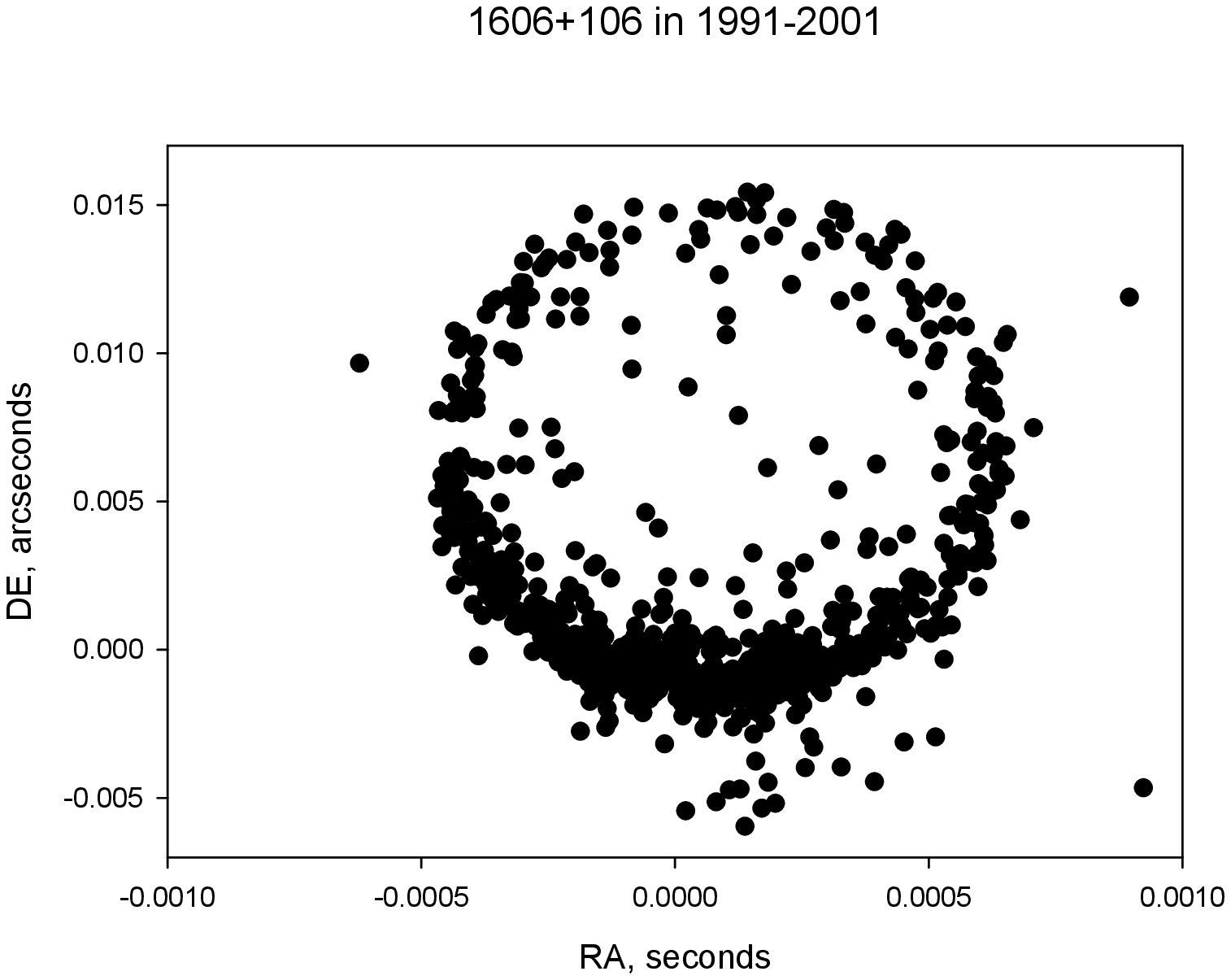}
		\end{center}
	\end{minipage}
\hfill
	\begin{minipage}[h]{0.49\linewidth}
		\begin{center}
			\includegraphics[scale=0.5, trim= 60 270 80 260,clip]{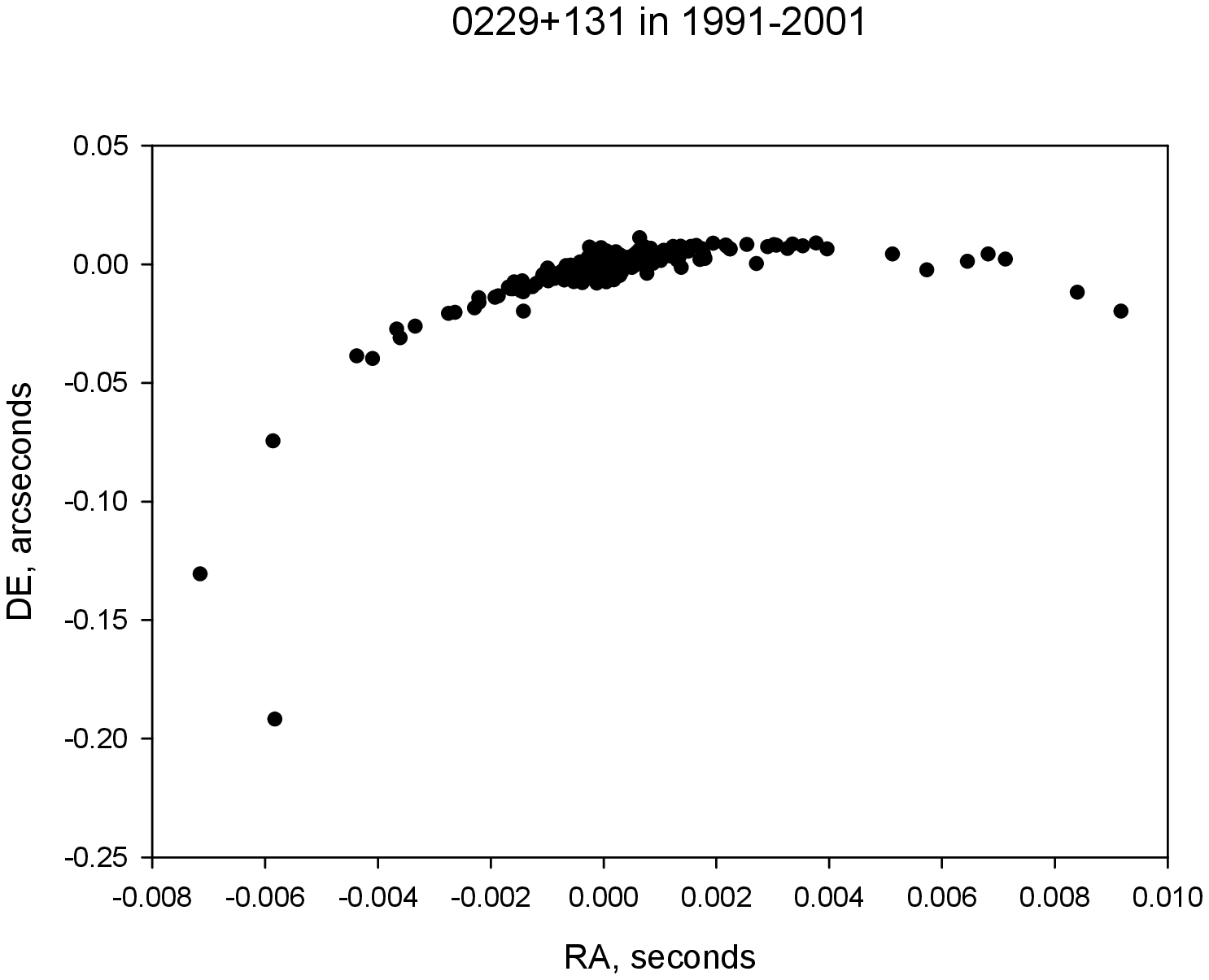}
		\end{center}
	\end{minipage}
\caption{The variations of the coordinates of the radio sources 1606+106 (left) and 0229+131 (right) in 1991-2001. A smaller minimum angle $\theta$ for 0229+131 results in larger size of the circle.}\label{fig:circ}
\end{figure}

We used the Least Squares collocation method to estimate the wet troposphere fluctuations with the OCCAM software. The mutual correlation between observables are introduced in the data adjustment process. The difference between the VLBI estimates of the wet troposphere delays and independent radiometer data is typically within 3-6 mm, or 10-20 ps (Titov \& Stanford, 2013). Thus, the impact of the wet troposphere delay on the astrometric light deflection angle estimated near the Sun is negligible. 
\begin{figure}[h]
	\begin{minipage}[h]{0.49\linewidth}
		\begin{center}
			\includegraphics[scale=0.5, trim= 60 270 80 260,clip]{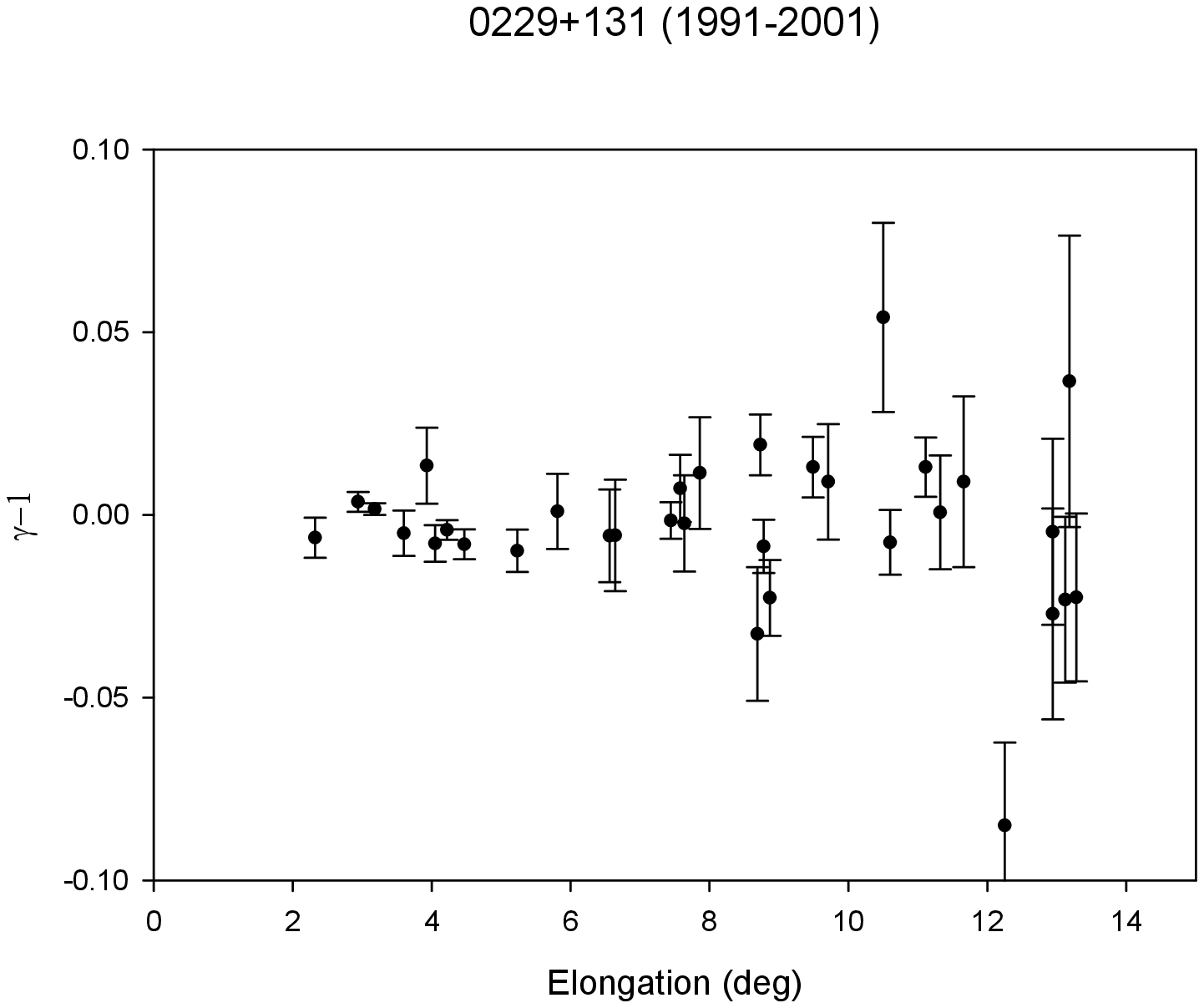}
		\end{center}
	\end{minipage}
\hfill
	\begin{minipage}[h]{0.49\linewidth}
		\begin{center}
			\includegraphics[scale=0.5, trim= 60 270 80 260,clip]{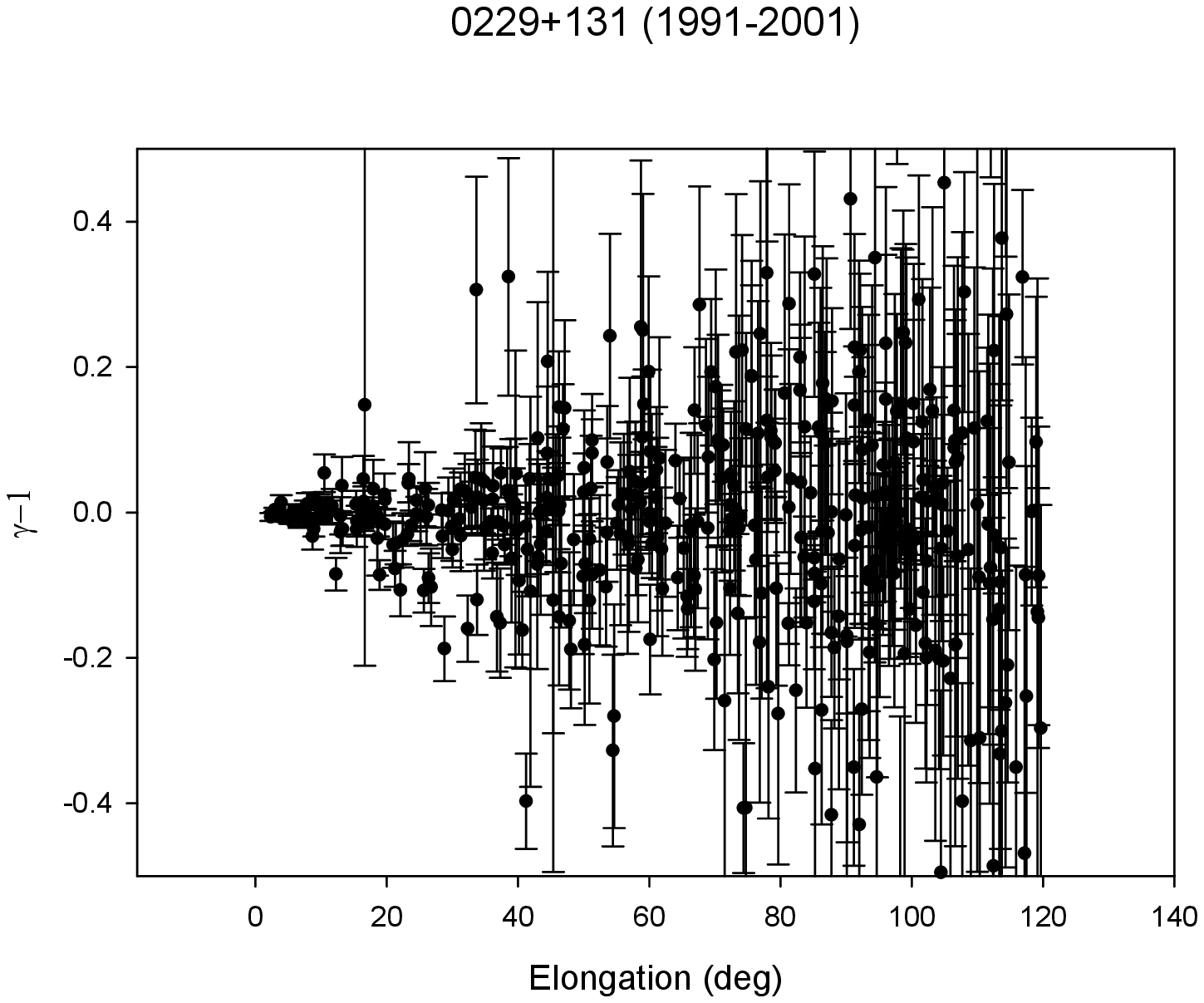}
		\end{center}
	\end{minipage}
\caption{Variations of the daily estimates of the PPN parameter $\gamma$ for radio source 0229+131 in 1991-2001 at the different scale of angle $\theta$.}\label{fig:circ}
\end{figure}

Fig 2 shows the variations of the coordinates of two selected radio source over 10 years (1991-2001). Each point on the plot corresponds to a daily position estimated from a single 24-hour session. About 10 annual circles are drawn by each radio source on the plots. The catalogue (non-deflected) positions are at the reference origin, and the total deflection angle is equal to distance to the point from the reference origin. The ecliptic latitude of the four sources are different, therefore, elongation angles vary over different ranges. As a result, the deflection angle does not change essentially for the former radio sources (the reference origin is in the middle of the ring) with a magnitude about 4 mas. In contrast, for the latter case, the deflection angle varies from zero to 0$''$.2 (for the minimum elongation of 3$^\circ$), and the reference origin lies near the edge of the ring.  

The variations of the deflection angle are easily converted with formula (1) to the variations of the PPN parameter $\gamma$. Fig 3 shows the variations of the deflection angle as a function of $\theta$ for the reference radio source 0229+131 at different scales. This radio source is almost on the ecliptic plane (ecliptic latitude $\epsilon\approx$ -1$^\circ$.5), and the elongation angle is limited by the technical capability of observing near the Sun. On 29 April, 1997 (session code 97APR29XE) the mean elongation angle was 1$^\circ$.55, and the measured deflection reached a maximum among all selected radio sources, $\alpha$ = 0$''$.3100 +/- 0$''$.0009. However, this estimate was obtained from only 7 observations. The best accuracy was achieved on 1 May, 1996 (session code 96MAY01XO) with 122 observations. The deflection angle estimate is $\alpha$ = 0$''$.1588 +/- 0$''$.0002, and the relative accuracy is 0.0012, in spite of a larger mean elongation angle for this date. 

\vspace*{0.7cm}

\noindent {\large 4. CONCLUSION}

\smallskip

The effects of general relativity are explicitly contained in both components of the total VLBI delay model -- 
the gravitational and geometric delays. While the former component uses the individual barycentre positions of the 
radio telescopes to calculate the effect, the latter component is expressed in terms of the baseline between the radio telescopes. 
Coupling between both parts has not been investigated until now.

The Shapiro effect as measured by radars, and the deflection of light measured with traditional astronomical instruments are 
considered two independent tests of general relativity. In this paper we show that the total group delay model joins the two 
tests within one observational technique - geodetic VLBI. The gravitational delay that traditionally originated from the Shapiro 
effect is linked to the light deflection angle. Therefore, the two approaches, VLBI delay and angular, are absolutely equivalent.

For almost all realistic situations this angle does not depend on the baseline length, thus, a standard geodetic VLBI interferometer 
acts as a traditional astronomical instrument. In addition, the coordinate term explicitly presented in the conventional geometric delay 
model ceases to exist because it is compensated by the same effect in the gravitational delay with the opposite sign. 
Thus, the proposed alternative version of the general relativity contribution to the total VLBI group delay model is free from 
coordinate effects. Therefore, the two approaches, time delay and angular deflection, are absolutely equivalent.

The final equation of the general relativity contribution also comprises two smaller terms which are significant at 
very small angular separation between the deflecting body and distant radio source. These terms may be considered as 
an increment in the light deflection angle due to the additional time delay for propagation of the light from station 1 to station 2. 
This effect become significant at $\frac{b}{R}>0.1$.

\vspace*{0.7cm}

\noindent \textsc{\large acknowledgements}

\smallskip

This paper is published with the permission of the CEO, Geoscience Australia.

\vspace*{0.7cm}

\noindent {\large 5. REFERENCES}

{

\leftskip=5mm
\parindent=-5mm

\smallskip

Einstein, A., 1916, Annalen der Physik, 354, 769.

Eubanks, T.M., Carter, M.S., Josties, F.J., Matsakis,~D.N., McCarthy,~D.D., 1991, IAU Colloq.~127: Reference Systems, 256.

Klioner, S.A., 1991, In: Proc. of AGU Chapman Conference on Geodetic VLBI: Monitoring Global Change, Washington DC, 188.

Kopeikin, S.M., 1990, Sov. Astron., 34, 5.

Ma, C., Arias, E.F., Bianco, G., et al., 2009, IERS Technical Note, 35. 

Schuh, H., Behrend, D., 2012, J. Geodyn., 61, 68.

Shapiro, I.I., 1964, Phys. Rev. Lett., 13, 789.

Shapiro, I.I., 1967, Science, 157, 806.

Soffel, M.H., Wu, X., Xu, C.,  Mueller, J., 1991, \aj, 101, 2306.

Titov, O., Girdiuk, A.\ 2015, \aa, 574, AA128 

Titov, O., Stanford, L.\ 2013, 21st Meeting of EVGA, Reports of the Finnish Geodetic Institute, p.~151-154

Titov, O., Tesmer, V., Boehm, J.\ 2004, IVS for Geodesy and Astrometry 2004 GM Proceedings, 267 

Ward, W.R., 1970, \apj, 162, 345.

}

\end{document}